\documentclass[runningheads]{llncs}
\usepackage[T1]{fontenc}
\usepackage{newtxtext}       %
\usepackage[varvw]{newtxmath}       

\usepackage{graphicx}
\usepackage{multirow}
\usepackage{capt-of}
\begin{document}
\title{A-GE: Privacy-Preserving Age Assurance via Fine-Grained Motion}
%
%
\author{Mohammed Gharib\inst{1}\orcidID{0000-0001-9611-1693} \and
Giovanni Cascone\inst{1}\orcidID{0009-0007-0029-575X} \and
Martin Zizi\inst{1}\orcidID{0000-0003-2021-9328}}
\authorrunning{M. Gharib et al.}
%
\institute{Aerendir Mobile Inc., Mountain View, CA 94070, USA \\
\email{\{gharib,martin\}@aermob.com} , \email{giovanni@aermobguest.com}
}
\maketitle              
\begin{abstract}
Age assurance faces a fundamental privacy dilemma: meaningful assurance typically requires revealing identity or other sensitive information, whereas privacy-preserving alternatives often provide little assurance. Existing approaches either disclose sensitive information directly or shift the disclosure and trust requirement to an identity provider, while self-declared age provides little meaningful assurance. Thus, a fundamental gap remains between establishing an age property and revealing information beyond what is necessary to establish that property.
We present \textit{A-GE} (Age-Grounded Evidence), a privacy-preserving age assurance system that derives age-related evidence from fine-grained motion captured by commodity inertial measurement units (IMUs). A-GE exploits subtle motion dynamics that we hypothesize reflect underlying neuromuscular and physiological processes during ordinary device interaction. It requires neither identity information nor government identification and operates across available sensor sampling rates, including the approximately 60~Hz limit of web-based sensing, with configurable minimum-age thresholds ranging from 14 to 21 years.
We evaluate A-GE on more than 600 participants, with separate participant-disjoint cohorts for model development and final evaluation. At a 16-year minimum-age threshold, A-GE achieves up to 93.8\% accuracy with a comparable F1 score across the evaluated configurations. We further demonstrate client-side deployment on Android, iOS, and the web, and examine unintended attribute inference and residual privacy risks, characterizing both the privacy benefits and limitations of deriving age evidence from commodity motion.

\keywords{Privacy-preserving age assurance \and physiological motion \and IMU.}
\end{abstract}
\section{Introduction}
\label{sec:introduction}

Age assurance is increasingly required for access to age-restricted goods, services, and online content. Yet in many settings, the property that must be established is remarkably narrow: whether a user satisfies a specified age threshold, such as 18 years. Establishing this predicate, however, often requires substantially more information than the predicate itself. Presenting a government-issued identity document, for example, may disclose a user's name, date of birth, photograph, address, identification number, and other attributes irrelevant to the age decision. This mismatch between the information required for age assurance and the information disclosed to obtain it creates a fundamental privacy problem~\cite{icoAgeAssurance2024,opc2026privacyAgeAssurance}. 
Existing age-assurance mechanisms expose this tension in different ways. Traditional approaches rely on government-issued identification or other credentials, while online services may request photographs, facial images, or biometric information for age estimation. Although such mechanisms can provide meaningful assurance, they require collecting information that may be sensitive or identifying and may introduce risks of retention, linkage, profiling, or secondary use~\cite{opc2026privacyAgeAssurance,icoAgeAssurance2024}. Avoiding such disclosure, however, often reduces assurance: self-declared age imposes little privacy cost but provides little evidence that the declaration is truthful~\cite{icoAgeAssurance2024,shaffique2024mapping}. Thus, existing approaches commonly trade between meaningful assurance with additional disclosure and minimal disclosure with weak assurance. The challenge is to establish an age-related property while minimizing the information disclosed to the relying party.

Digital identity and credential systems provide an important improvement to this tradeoff. A user can obtain a digitally verifiable credential and disclose only an age-related claim to a relying party, substantially reducing disclosure at the point of use~\cite{shaffique2024mapping,opc2026privacyAgeAssurance}. This approach, however, introduces a different trust dependency: an issuer or identity provider must establish the user's underlying attributes before issuing the credential. The disclosure burden is therefore reduced for the relying party but shifted toward the credential infrastructure. This leaves an open question: \emph{can useful age assurance be obtained without requiring either direct disclosure of identity information or an identity-provider-mediated credential?}

We explore this question with \textit{A-GE}, a privacy-preserving age assurance system that derives age-related evidence from fine-grained motion captured during ordinary device interaction. A-GE exploits subtle, involuntary motion dynamics that we hypothesize partially reflect underlying physiological and neuromuscular processes and contain information that can be learned to establish whether a user satisfies a specified age threshold. This hypothesis is supported by emerging evidence that fine-grained inertial dynamics capture information about human interaction beyond intentional device movement; in particular, prior work has shown that related motion dynamics can provide evidence of human presence and distinguish live human interaction from automated or device-mediated activity~\cite{alive}. A-GE investigates the complementary question of whether these dynamics also contain sufficient information to establish an age-related property, without requiring a government-issued document, facial image, or identity credential.

A-GE is designed to operate across the sensing capabilities provided by the underlying platform. Although higher sampling rates can provide finer temporal resolution, A-GE remains effective at approximately $60$~Hz, the sampling regime available to web applications, while supporting higher rates on platforms that provide them. This enables the same sensing approach to operate across native mobile and browser-based environments without specialized hardware. A-GE further supports configurable minimum-age thresholds; we evaluate thresholds from 14 to 21 years to capture the variation in age limits across regulatory and application contexts~\cite{oecdAgeAssurance2025}.

We develop and evaluate A-GE with the participation of more than 600 contributors spanning diverse ages, genders, and demographic backgrounds, with participant-disjoint model development and evaluation. Approximately 250 participants contributed to the model development and cross-validation, while approximately 350 participants form a completely held-out test cohort that is never exposed to training and development. For a 16-year minimum-age threshold, A-GE achieves up to 93.8\% accuracy with a comparable F1 score under the evaluated configurations. We systematically evaluate observation duration, sampling rate, and minimum-age thresholds, characterizing the sensing requirements and generality of A-GE across practical deployment conditions. We further implement A-GE as native Android and iOS applications and as a client-side web system, demonstrating deployment across native and browser-based environments while keeping raw sensor measurements on the user's device.
The goal of A-GE is not to assume that motion measurements are inherently non-sensitive, but to reduce the information that must be disclosed to establish an age property. A-GE therefore derives age-related evidence locally from fine-grained motion rather than requiring identity-derived information, while our privacy analysis examines what additional information may remain inferable from the sensing modality. A-GE makes the following contributions:
\begin{itemize}

\item \textbf{Privacy-preserving age assurance from fine-grained motion.}
We introduce A-GE, and demonstrate that this previously underexplored sensing signal
contains information sufficient to support practical age assurance.

\item \textbf{Configurable and platform-independent age assurance.}
A-GE supports application-specific minimum-age thresholds and operates across sensing capabilities provided by different platforms, including the approximately $60$~Hz available to web browsers.

\item \textbf{Participant-disjoint empirical characterization.}
More than 600 contributors participated in the development and the evaluation of A-GE models on a completely held-out cohort that is excluded from model training,
development, and selection. We systematically characterize the effects of observation duration, sampling rate, and age threshold on performance.

\item \textbf{Privacy analysis of motion-based age assurance.}
We investigate whether motion measurements used for age assurance expose attributes beyond age, and characterize the
privacy limitations and residual risks. 

\end{itemize}

The remainder of the paper is organized as follows. Section~\ref{sec:problem} formalizes the age-assurance problem and the adversary model. Section~\ref{sec:background} reviews related work on age assurance. Section~\ref{sec:design} presents A-GE's design. Section~\ref{sec:methodology} describes the  evaluation protocol and metrics. Section~\ref{sec:privacy-analysis} examines data minimization, unintended inference, linkability, and residual privacy risks, while Section~\ref{sec:evaluation} evaluates performance across a wide range of conditions. Section~\ref{sec:conclusion} concludes the paper.

\section{Problem Statement and  Adversary Model}
\label{sec:problem}


\textit{\textbf{Problem Statement and Requirements.}}
A-GE aims to enable a relying application to establish an age-related predicate about a user while minimizing the information required from that user. Specifically, for an application-defined threshold $T$, A-GE seeks to establish
    $\mathsf{Age}(u) \geq T$,
without recovering the user's identity or requiring disclosure of identity-derived attributes such as a name, date of birth, or government identification. The resulting evidence is probabilistic and is therefore not intended to replace authoritative, identity-backed age verification where deterministic or legally binding verification is required. Instead, A-GE targets applications where useful age assurance, minimal identity disclosure, and low user friction are the primary objectives. This scope leads to the following requirements.
\noindent{\textit{i) Age assurance and data minimization:}}
A-GE should provide meaningful evidence for the requested age predicate without relying on self-declaration or requiring unnecessary identity disclosure. During operation, raw sensor measurements remain on the user's device and only the resulting age-assurance outcome is disclosed.
\noindent{\textit{ii) Frictionless and commodity-device operation:}}
A-GE should require no identity document, facial image, specialized physiological sensor, or explicit verification procedure beyond ordinary interaction with a commodity device, and should operate using sensing capabilities already available on commodity devices.
\noindent{\textit{iii) Configurable age predicates:}}
The same sensing and inference framework should support application-defined minimum-age thresholds rather than being tied to a single definition of adulthood. This allows the mechanism to be configured for different age thresholds.

\textit{\textbf{Adversary Model and Privacy Goals.}}
A-GE is designed primarily to minimize information disclosure rather than to provide a comprehensive defense against a compromised sensing or execution environment. We consider an adversary that may observe the age-assurance outcome, know the A-GE architecture and inference procedure, and attempt to infer information about the user beyond the age property required by the relying application. The adversary may also possess knowledge of the sensing modality and access to sensor observations from other users.\\
\noindent{\textit{Sensing and execution integrity.}}
We assume that the sensor measurements supplied to A-GE originate from the device's physical sensing stack and that an adversary cannot arbitrarily replace, inject, or modify IMU samples at the input of the inference pipeline. Compromise of the operating system, sensor drivers, application runtime, or other components providing this sensing and execution path is outside our scope. Without this assumption, an adversary could trivially supply a prerecorded observation from the target age class and bypass any sensor-based inference mechanism. Accordingly, A-GE establishes an age-related property of the physical sensor observation rather than of an arbitrary byte sequence presented to the classifier. Attacks that compromise the client execution environment or alter the age-assurance outcome after local inference are outside the scope of this work.\\
\noindent{\textit{Privacy goals.}}
A-GE's primary privacy objective is to establish the application's requested age predicate without requiring disclosure of identity-derived information. In particular, the process should not require a government-issued identity document, exact date of birth, facial image, or other identifying attributes solely to establish the age predicate. During operation, raw IMU measurements remain on the user's device and only the resulting age-assurance outcome is disclosed.
The privacy objective does not imply that IMU measurements are intrinsically non-sensitive. Fine-grained motion may contain information about characteristics beyond age, including behavioral or physiological properties. A-GE therefore follows a limited-purpose inference principle: the system is trained and operated to establish an application-defined age predicate rather than to identify the user or intentionally infer unrelated attributes.

\section{Background and Related Work}
\label{sec:background}
This section provides the background and related work underlying A-GE. We first describe the physiological basis motivating the use of fine-grained inertial motion, and then review existing age-assurance approaches and position A-GE within this landscape.
\paragraph{\textbf{Physiological Basis of IMU Motion:}}
A-GE is motivated by the hypothesis that fine-grained variations in accelerometer and gyroscope measurements during ordinary device interaction may reflect subtle human motor dynamics. Prior work has shown that related fine-grained IMU dynamics can provide evidence of human presence and distinguish live human interaction from automated or device-mediated activity~\cite{alive}. Human movement arises from neural and neuromuscular processes that produce mechanical forces and small physical movements, including physiological tremor that can be captured by accelerometers~\cite{cortico2015,tremor}. When a smartphone or wearable is held or worn, such movements can become coupled to the device and appear in its inertial measurements alongside intentional motion, environmental disturbances, and sensor noise. The hypothesized causal pathway is therefore\\
\small{
$\mathrm{Neural\ Activity} \rightarrow \mathrm{Neuromuscular\ Activation} \rightarrow\mathrm{Fine\text{-}Grained\ Motion} 
    \rightarrow  \mathrm{IMU\ Measurements}$}, rather than a direct relationship between neural activity and IMU measurements. We refer to this hypothesized component as \emph{neuromuscular micro-motion}; the term does not imply that an IMU directly measures neural activity. Conventional inertial-sensing pipelines commonly apply filtering and signal separation to suppress noise, gravity, or components considered irrelevant to the target activity~\cite{imuLiveness,uciHAR}. In contrast, A-GE does not explicitly filter or isolate a physiologically defined micro-motion component; it operates on the sensor measurements without assuming which signal components carry age-related information.
To provide preliminary evidence consistent with this physiological hypothesis, we conducted an exploratory cross-modal study with synchronized IMU and EEG measurements from six participants, using E-MOTIV Earpods for EEG acquisition. We applied principal component analysis to the IMU measurements, isolated the 4--8~Hz component, and compared its temporal energy with EEG activity in the corresponding frequency range. The resulting measures exhibited a mean correlation of approximately $0.19$. This observation is exploratory and does not establish causality or imply that the IMU measures neural activity. Rather, together with prior evidence that accelerometer measurements can capture physiological tremor and motor-related kinematic dynamics~\cite{cortico2015,tremor}, it provides preliminary support for the hypothesis that fine-grained inertial measurements can contain information arising from human motor dynamics. A-GE subsequently tests whether such information can be exploited for age assurance, without requiring a specific physiological mechanism to be established.

\paragraph{\textbf{Related Work:}}
Age assurance encompasses techniques for establishing whether an individual satisfies an age-related eligibility requirement. Existing approaches range from self-declaration and document-based verification to digital credentials, biometric estimation, behavioral profiling, and emerging sensing-based techniques~\cite{shaffique2024mapping,ukAgeAssuranceDataStudy2026}. Although these approaches differ substantially in mechanism and assurance strength, they can be organized from a privacy perspective into two broad categories. \emph{Identity- and credential-based approaches} establish age from information or credentials associated with an authoritative identity, whereas \emph{sensing- and behavior-based approaches} infer age from observable characteristics of the user or their interaction. This distinction is useful because the two categories address the privacy problem differently: the former can minimize what is disclosed to the relying party but generally requires an identity or trusted issuer to establish the underlying age, while the latter can avoid identity disclosure but may require collecting sensitive biometric, behavioral, or sensor information.

\textit{Identity- and credential-based age assurance.}
Self-declaration represents the simplest approach. It requires little or no personal information and is easy to deploy, but provides essentially no evidence that the claimed age is correct~\cite{shaffique2024mapping}. Government-issued identification and other authoritative credentials provide substantially stronger evidence, but may disclose considerably more information than is necessary for an age predicate. A relying party that only needs to establish whether a user is at least 18 years old does not inherently need the user's name, exact date of birth, address, photograph, or identification number. Nevertheless, conventional verification mechanisms may expose some or all of these attributes~\cite{opc2026privacyAgeAssurance,icoAgeAssurance2024}.
Privacy-preserving credential systems reduce this disclosure by separating the initial establishment of age from the subsequent presentation of an age claim. Anonymous credentials~\cite{anonymCred}, selective-disclosure credentials~\cite{selectiveDisclosCred}, verifiable credentials~\cite{verifiableCred}, and zero-knowledge proofs~\cite{ZK} provide cryptographic mechanisms through which a user can demonstrate an age-related predicate such as $\mathsf{age}(u)\geq T$ without repeatedly presenting the underlying identity document. These techniques can substantially reduce the information available to a relying party~\cite{selectiveDisclosCred,ZK}, and recent work has explored their application to privacy-preserving age assurance and existing digital credentials~\cite{PPAge,crescent}.
These systems, however, shift rather than eliminate the underlying trust and disclosure requirements. Before an age credential can be presented, an issuer must establish the user's age using an authoritative source, commonly an identity document or trusted identity record. The relying party may consequently receive only a minimal age claim, but the user must still trust the issuer or identity infrastructure with the information required to establish that claim. Credential systems also introduce dependencies on issuance, storage, revocation, interoperability, and trust frameworks~\cite{verifiableCred}. Thus, credential-based approaches primarily minimize disclosure during the \emph{subsequent assertion} of age; A-GE instead investigates whether the age property can be established without the initial identity-based age-establishment step.

\textit{Sensing- and behavior-based age assurance.}
An alternative is to infer age directly from observable characteristics rather than from identity information. Biometric and camera-based approaches estimate age from characteristics such as facial appearance~\cite{faceAgeSurvey}, voice~\cite{voiceAge}, and other physiological attributes~\cite{faceVoiceAge,faceAgeSurvey2}. These methods can avoid direct presentation of an identity document, but the inference process itself requires acquisition and processing of sensitive biometric information. Facial approaches, for example, may generate biometric representations in addition to the final age estimate, creating privacy risks associated with collection, retention, linkage, and secondary use~\cite{biometricPrivacy}. The privacy properties of such systems therefore depend not only on the age result disclosed to the relying party but also on the information exposed during inference.

Behavioral approaches infer age-related characteristics from online behavior, interaction patterns, or other observable user activity~\cite{behaviourAge,eu2024mappingage}. Although they can avoid explicit identity documents, they typically require observations over time to construct a behavioral profile, introducing opportunities for persistent profiling and secondary use. Moreover, because the relationship between behavior and age is indirect, the resulting assurance depends on the stability and representativeness of the observed behavior.

Motion sensing provides another form of direct observation. Prior work has demonstrated that inertial measurements can contain age-related information, including through gait-based age prediction~\cite{gait} and the detection of age-related differences in walking~\cite{imuAge}. These studies establish the feasibility of extracting age information from IMU signals, but rely on structured locomotion, typically requiring the user to walk while carrying or wearing an inertial sensor. A-GE instead investigates whether age-related information is present in fine-grained motion generated during ordinary interaction with a handheld device, without requiring a prescribed activity or gait measurement. This distinction also has privacy implications: inertial measurements need not contain explicit identity information to reveal characteristics of the person generating them~\cite{gait}. A-GE therefore treats the sensing modality itself as part of the privacy analysis, while avoiding the identity-establishment dependency of credential-based approaches and the persistent observation required by behavioral profiling. 
The distinction between these approaches is summarized in Table~\ref{tab:related-work}. Rather than treating privacy as the simple absence of an identity document, we compare approaches across the properties most relevant to privacy-preserving age assurance: the strength of age evidence, identity disclosure, third-party trust, configurable age predicates, data minimization, disclosure of raw sensing data, and user friction. 

\begin{table*}[t]
\centering
\caption{Comparison of age-assurance approaches. `\checkmark'' denotes supported, `--'' unsupported, and `Partial'' conditional support; }
\label{tab:related-work}
\small
\setlength{\tabcolsep}{4pt}
\resizebox{\linewidth}{!}{
\begin{tabular}{llccccccc}
\hline
\textbf{Category} &
\textbf{Approach} &
\textbf{No ID} &
\textbf{No 3rd} &
\textbf{Configurable} &
\textbf{Data} &
\textbf{No Raw} &
\textbf{Frictionless} &
\textbf{Age} \\
\textbf{} &
\textbf{} &
\textbf{} &
\textbf{Party Trust} &
\textbf{Threshold} &
\textbf{Minimization} &
\textbf{Data} &
\textbf{} &
\textbf{Evidence} \\
\hline
\multirow{3}{1in}{\textit{Identity- and credential-based}}
&Self-declaration~\cite{shaffique2024mapping}& Low & \checkmark & \checkmark & \checkmark & \checkmark & \checkmark & --  \\
&Government ID~\cite{icoAgeAssurance2024}& High
 & -- & \checkmark & Partial & -- & -- & --  \\
&Age credentials~\cite{PPAge} & High & \checkmark & -- & \checkmark & \checkmark & \checkmark & -- \\
\hline
\multirow{5}{1in}{\textit{Biometric, behavioral, and motion-based}}
&Facial age estimation~\cite{faceAgeSurvey}  & High& \checkmark & \checkmark & \checkmark & -- & Partial & \checkmark \\
&Voice age estimation~\cite{voiceAge}& Partial & \checkmark & \checkmark & \checkmark & -- & Partial & \checkmark  \\
&Behavioral profiling~\cite{behaviourAge}& Partial & \checkmark & \checkmark & \checkmark & -- & \checkmark & \checkmark  \\
&Motion sensing~\cite{gait}& Partial & \checkmark & \checkmark & \checkmark & Partial & Partial & Partial  \\
\hline
&\textbf{A-GE (this work)}
 & $>93\%$ & \checkmark & \checkmark & \checkmark & \checkmark & \checkmark & \checkmark  \\
\hline
\end{tabular}}
\end{table*}

\section{A-GE Design}
\label{sec:design}

\begin{figure*}[!b]
    \centering         
    \includegraphics[width=\linewidth]{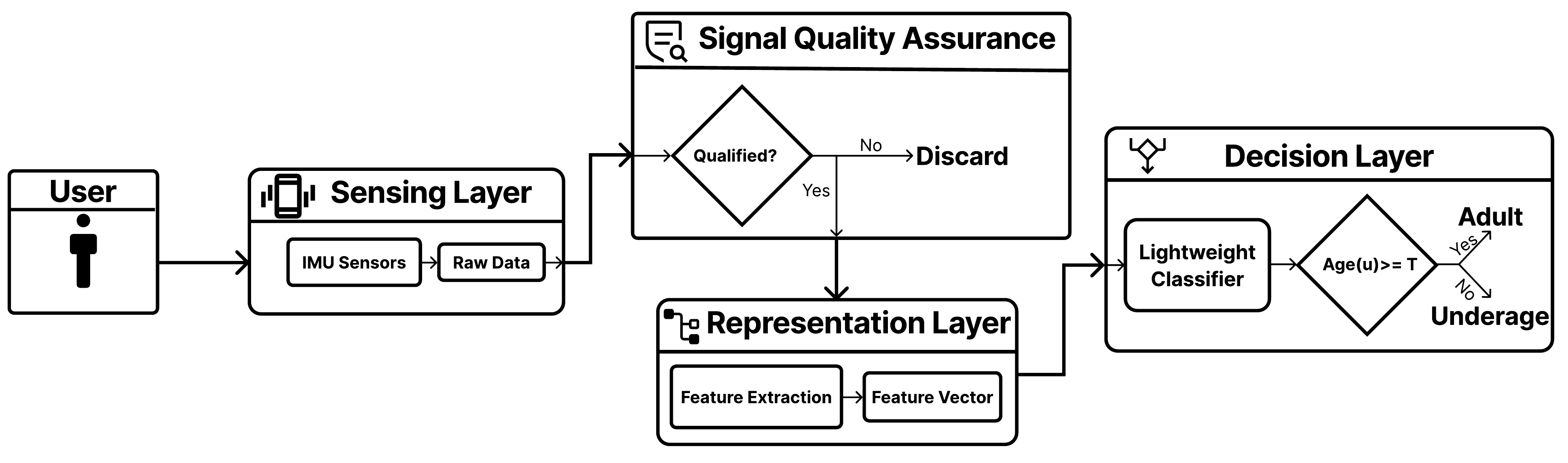}
    \caption{Overview of the A-GE Framework }
    \label{fig:system-overview}
\end{figure*}

Figure~\ref{fig:system-overview} illustrates A-GE architecture, which comprises five logical layers: \emph{User}, \emph{Sensing}, \emph{Signal Quality Assurance}, \emph{Representation}, and \emph{Decision}. The user interacts with a commodity smartphone or wearable, while the remaining layers operate locally on the device. The resulting age-assurance outcome, rather than the underlying sensor measurements, is exposed to the relying application. 
The processing pipeline is
$X
\xrightarrow{\mathsf{SQA}}
X'
\xrightarrow{\mathsf{Rep}}
\mathbf{z}
\xrightarrow{\mathsf{Dec}}
\hat{y}$,
where $X$ denotes the acquired inertial measurements, $\mathsf{SQA}$ performs signal-quality assurance, $\mathbf{z}$ denotes the resulting feature representation, and $\mathsf{Dec}$ produces the age-assurance outcome $\hat{y}$. The \emph{Sensing Layer} acquires accelerometer and gyroscope measurements; the \emph{Signal Quality Assurance Layer} determines whether an observation window is suitable for inference; the \emph{Representation Layer} transforms accepted measurements into statistical and spectral descriptors; and the \emph{Decision Layer} maps the resulting representation to the application-specific age predicate. 

\paragraph{\textbf{Sensing and Data Acquisition.}}
A-GE uses commodity inertial measurement units (IMUs), specifically
accelerometers and gyroscopes available on modern smartphones and wearable
devices. No specialized sensing hardware is required. The system operates
over a broad range of sensor sampling rates, including approximately
$60$~Hz and higher sampling rates available through native mobile
interfaces.
For each observation, the sensing layer collects a finite temporal window of
three-axis accelerometer and gyroscope measurements. Let
    $X =
    \left\{
    \mathbf{a}(t),\mathbf{g}(t)
    \right\}_{t=1}^{N}$,
where
    $\mathbf{a}(t) =
    [a_x(t),a_y(t),a_z(t)]$
and
    $\mathbf{g}(t) =
    [g_x(t),g_y(t),g_z(t)]$.
The number of samples $N$ depends on the effective sensor sampling rate and
the observation duration.
A-GE does not require a fixed sampling rate, enabling deployment across heterogeneous devices and application environments. In particular, the approximately $60$~Hz regime available through browser-based sensor interfaces is sufficient for effective inference, allowing A-GE to operate in web environments without specialized native sensing access. We empirically evaluate the effects of both observation duration and sampling rate in Section~\ref{sec:evaluation}.

\paragraph{\textbf{Signal Quality Assurance.}}
Raw inertial measurements are affected by the physical state of the device
and by the quality of the interaction during acquisition. A sensor window
may contain too few samples, may correspond to a stationary device rather
than an actively held device, or may contain motion artifacts sufficiently
large to obscure the fine-grained signal used by A-GE. Examples include a
smartphone resting on a desk, abrupt or excessive device motion, and
forceful tapping or other impulsive interactions with the device.
A-GE therefore introduces an explicit Signal Quality Assurance (SQA) layer
between sensing and representation. Rather than attempting to remove every
potential artifact through filtering, SQA determines whether a window is
appropriate for inference. The objective is not to transform a poor-quality
signal into an apparently valid one, but to prevent an unreliable observation
from influencing the age decision.

The SQA layer evaluates properties of the acquired window including
\emph{sampling completeness}, \emph{physical interaction state}, and
\emph{motion quality}. Sampling completeness verifies that the number and
temporal distribution of received sensor samples are consistent with the
expected acquisition conditions. The interaction checks distinguish
meaningful device interaction from conditions such as a phone remaining
stationary on a surface. Motion-quality checks identify windows containing
excessive or abrupt motion, including strong impulses such as harsh tapping,
that can dominate the inertial signal and obscure the fine-grained dynamics
used by the subsequent representation.
If a window fails the quality criteria, A-GE rejects the window and does not
perform age inference on it. This rejection mechanism is deliberately
preferable to aggressive artifact removal because filtering can alter or
eliminate signal components that may contain information relevant to age.
Thus, SQA acts as a gate on the validity of the input rather than as a
conventional denoising stage.
It establishes whether a particular
observation contains sufficient signal quality for the model to produce a
meaningful age-assurance result.

\paragraph{\textbf{Age-Related Feature Representation.}}
For accepted windows, A-GE computes a feature representation that captures complementary statistical, temporal, and frequency-domain characteristics of the inertial measurements. Rather than relying on a single manually selected signal component, the representation combines multiple feature families:
\begin{itemize}
    \item \textbf{Distributional statistics.}
    Percentiles, mean, standard deviation, interquartile range, and related
    statistics characterize the amplitude distribution of the inertial
    signal.

    \item \textbf{Higher-order statistical moments.}
    Skewness and kurtosis capture asymmetry and tail behavior that are not
    represented by first- and second-order statistics alone.

    \item \textbf{Temporal-domain dynamics.}
    Zero-crossing rates and related temporal descriptors characterize the
    fine-grained variation and oscillatory behavior of sensor
    axes.

    \item \textbf{Frequency-domain characteristics.}
    Spectral descriptors, including dominant-frequency characteristics,
    capture periodic and oscillatory structure in the inertial signal.

    \item \textbf{Magnitude-based descriptors.}
    Features computed from signal magnitude characterize
    aggregated motion while reducing dependence on an individual axis.
\end{itemize}
While proprietary constraints prevent sharing the exhaustive feature list, these disclosed families and examples provide sufficient methodological detail for researchers to fully replicate the model’s design space and signal representation.
Importantly, the representation is not intended to identify a particular physiological mechanism by assigning individual features to specific neuromuscular processes. Rather, it provides a data-driven representation of the fine-grained inertial dynamics from which the age-assurance model learns age-related structure.

\paragraph{\textbf{Age Assurance Model.}}
The final A-GE decision is produced by a supervised classification model
operating on the feature representation extracted from an accepted sensor
window. We use Extreme Gradient Boosting (XGBoost), a gradient-boosted
decision-tree model, as the decision mechanism. XGBoost is well suited to
the heterogeneous feature representation used by A-GE as it can model
nonlinear relationships and interactions among statistical and spectral
descriptors without requiring the features to share a common scale or
distribution.
Let $\mathbf{z}$ denote the feature vector extracted from an accepted
observation and let $T$ denote the application-specific minimum-age
threshold. The decision layer computes
    $\hat{y}
    =
    f_{\theta}(\mathbf{z};T),$
where $f_{\theta}$ is the trained A-GE model and $\hat{y}$ indicates whether
the observation satisfies the configured age requirement. 
To train the model, observations from an individual participant are confined to a
single training, development, or test partition. This prevents the model
from exploiting participant-specific characteristics shared across multiple
recordings and ensures that the reported performance measures
generalization to previously unseen participants.
The A-GE decision model consists of 300 boosted decision trees with a maximum tree depth of 6 and a learning rate of 0.05. We use the binary
classification objective with log-loss as the evaluation metric. No subsampling, feature subsampling, or class weighting is applied. The predicted probability $p$ is converted to the final binary age-assurance decision using a threshold of
$0.5$. 

\section{Experimental Methodology}
\label{sec:methodology}

We evaluate A-GE using a participant-disjoint protocol, age thresholds from 14 to 21 years, and complementary performance and privacy analyses. Participants took part in our experiments by holding the smartphones during multiple continuous recording sessions using dedicated Android and iOS applications that monitored accelerometer and gyroscope measurements during ordinary handheld interaction. The logged participant attributes were limited to gender, exact age, and dominant hand; no names, government identification, addresses, facial images, or other identity attributes were obtained.
Participants were divided into two strictly separated cohorts. The development cohort contains 273 participants (178 male, 95 female), aged 4--83 years (median 21.5); 27 reported left-hand dominance and 246 right-hand dominance. The held-out evaluation cohort contains 359 participants (227 male, 132 female), aged 4--69 years (mean 28.7); 9 reported left-hand dominance and 350 right-hand dominance. Separation occurs before window segmentation, and the evaluation cohort is excluded from training, cross-validation, model selection, and all other development procedures. This prevents leakage from persistent individual characteristics, such as motor behavior and device interaction patterns, that could inflate generalization performance. Unless otherwise stated, experiments varying observation duration, sampling rate, or age threshold hold other conditions fixed.

We evaluate age thresholds from 14 to 21 years, using 16 years as the primary threshold. Labels are derived from recorded ages, and the development data are balanced between classes before training and cross-validation to limit class-distribution bias and facilitate comparisons across thresholds. The held-out cohort is evaluated both on a balanced subset with equal numbers of participants from each class and in its entirety under the natural age distribution. Accuracy and F1 scores are closely comparable between these conditions, suggesting that the principal findings are not artifacts of class balancing. Table~\ref{tab:threshold-demographics} reports the male/female composition of the two age classes at each threshold. 
Accuracy and F1 score are the primary binary age-assurance metrics, measuring overall correctness and the balance of precision and recall, respectively. Auxiliary experiments assess information potentially recoverable beyond the intended binary age property. Continuous age inference is evaluated using mean absolute error (MAE), root mean squared error (RMSE), coefficient of determination ($R^2$), Pearson correlation, and the fraction of predictions within specified age-error intervals. Gender and handedness inference are evaluated using ROC-AUC, accuracy, and F1 score. These auxiliary analyses characterize potential information leakage rather than intended A-GE capabilities. For probability-based analyses, $P(\mathrm{adult}\mid X)$ denotes the probability assigned to the adult class for observation $X$.

\begin{table}[t]
\centering
\begin{minipage}[t]{0.45\columnwidth}
    \centering
    \caption{Participant distribution across evaluated age thresholds.}
    \label{tab:threshold-demographics}
    \scriptsize
    \setlength{\tabcolsep}{3pt}
    \renewcommand{\arraystretch}{1.05}
    \begin{tabular}{c|cc|cc}
        \hline
        \multirow{2}{*}{Threshold} &
        \multicolumn{2}{c|}{Under Threshold} &
        \multicolumn{2}{c}{At/Above Threshold} \\
        & Male & Female & Male & Female \\
        \hline
        14 & 26 & 26 & 201 & 106 \\
        15 & 36 & 29 & 191 & 103 \\
        16 & 41 & 32 & 186 & 100 \\
        17 & 45 & 33 & 182 & 99 \\
        18 & 47 & 36 & 180 & 96 \\
        19 & 55 & 37 & 172 & 95 \\
        20 & 60 & 38 & 167 & 94 \\
        21 & 66 & 39 & 161 & 93 \\
        \hline
    \end{tabular}
\end{minipage}
\hfill
\begin{minipage}[t]{0.53\columnwidth}
    \centering
    \captionof{figure}{Age prediction error across five-year age buckets.}
    \label{fig:age-prediction}
    \includegraphics[width=\linewidth]{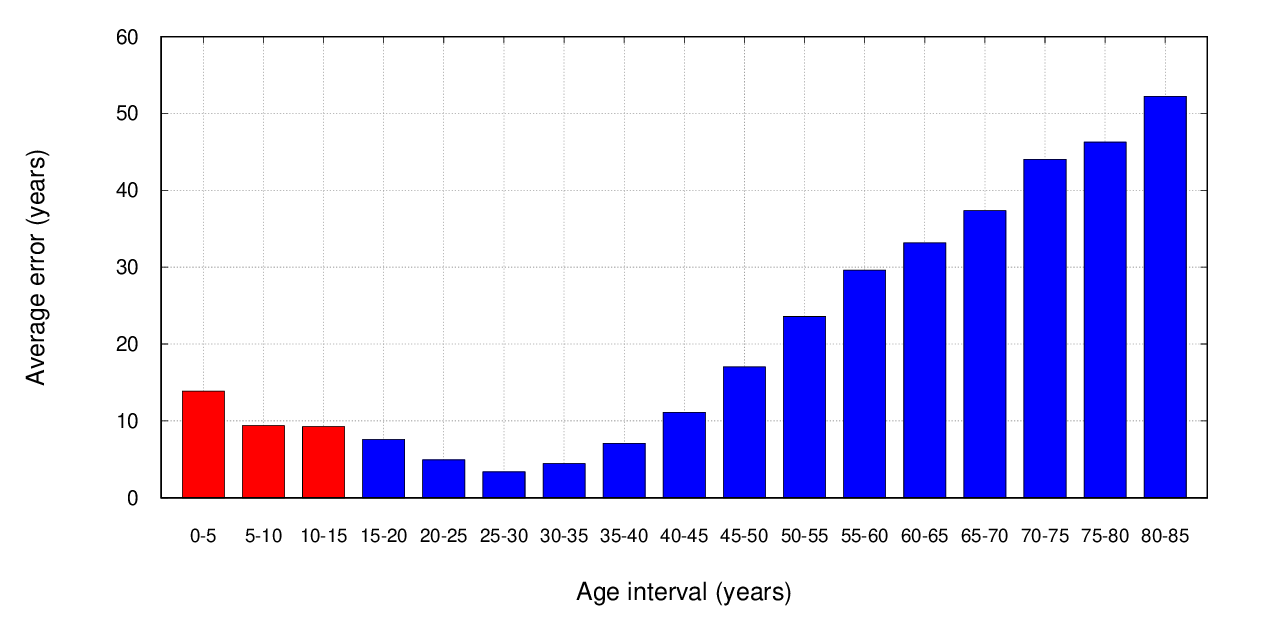}
\end{minipage}
\end{table}

\section{Privacy Analysis}
\label{sec:privacy-analysis}

Privacy is a primary design objective of A-GE rather than a consequence of omitting explicit identifiers. Although A-GE does not require a government identity document, facial image, name, date of birth, or identity-provider credential, its underlying IMU measurements are potentially informative. The key privacy question is therefore not only what A-GE discloses during age assurance, but also what additional information may be encoded in the sensing modality if the measurements were exposed. We investigate this question through auxiliary inference of exact age, gender, and handedness, and separately consider the potential for identity and cross-session linkability. While raw IMU measurements remain on the user's device in deployment, these experiments characterize the residual information content of the modality beyond the intended age predicate.
Specifically, we evaluate exact-age inference through regression, as well as gender and handedness inference. 
The objective is not to claim that the
tested attributes exhaust all possible information leakage, but rather to
provide empirical evidence about several plausible attributes that may be
encoded in fine-grained motion.

\textit{\textbf{Exact-age inference.}}
We first train a regression model to predict participants' exact ages from the same six-second motion observations used for age assurance, assessing whether the representation encodes information beyond A-GE's binary age threshold. On an evaluation set of 273 participants, the model achieves a mean absolute error (MAE) of $8.53$ years, an RMSE of $11.18$ years, $R^2=0.309$, and a Pearson correlation of $0.578$. Only $6.3\%$, $12.8\%$, $19.8\%$, and $33.5\%$ of predictions fall within one, two, three, and five years of actual age, respectively. Figure~\ref{fig:age-prediction} further illustrates prediction errors across five-year age buckets, revealing substantial variation across age ranges; estimates for the youngest and oldest buckets are less reliable due to limited participant counts. These results indicate that the IMU signal may leak some age information beyond A-GE's binary threshold, but large prediction errors limit its usefulness for age estimation.

\textit{\textbf{Gender inference.}}
We next evaluate whether the IMU representation supports inference of gender, which A-GE does not require. Using the same six-second IMU observations, a binary female/male classifier achieves $72.6\%$ accuracy, an F1 score of $72.2\%$, and an AUC of $0.807$, indicating that the modality contains some gender-related information. However, this inference remains imperfect, and the underlying IMU measurements are processed locally and never leave the device, limiting the exposure of this information.

\textit{\textbf{Handedness inference.}}
We similarly evaluate two forms of handedness inference. First, a classifier distinguishing dominant-hand from non-dominant-hand observations achieves $54.6\%$ accuracy, an F1 score of $54.5\%$, and an AUC of $0.601$. Second, a classifier distinguishing left-hand from right-hand observations achieves $56.8\%$ accuracy, an F1 score of $56.5\%$, and an AUC of $0.614$. Both results are close to the performance expected from a weakly informative classifier, particularly compared with the gender-inference results. The left-versus-right experiment may additionally capture differences in device orientation and phone positioning associated with holding the device in the left or right hand rather than a physiological property of the participant. The dominant-versus-non-dominant experiment is also affected by the composition of the collected dataset, which contains substantially fewer left-handed participants. Accordingly, these experiments provide only limited evidence of handedness leakage and should not be interpreted as establishing that A-GE can reliably infer handedness.

Accordingly, A-GE does not claim to eliminate privacy risks associated with
sensor-based inference. Its privacy contribution is instead to reduce the
information that must be disclosed to obtain age assurance: a user can
demonstrate an application-specific age property without transmitting an
identity document, facial image, or raw inertial measurements to the
relying party. The empirical privacy analysis makes the remaining
tradeoff explicit by showing both the utility of the modality for age
assurance and the extent to which the same modality supports selected
secondary inferences.

\section{Evaluation}
\label{sec:evaluation}

This section evaluates A-GE's performance and robustness across age-assurance and deployment conditions, following the protocol in Section~\ref{sec:methodology}. We establish a reference operating point with a $16$-year minimum-age threshold, a $5$-second observation window, and a $500$~Hz sampling rate, achieving $89\%$ accuracy and an F1 score of $88.7\%$ before temporal aggregation. This baseline provides a reference for assessing improvements rather than representing A-GE's maximum performance. We then vary individual factors to examine support for different age thresholds, the effects of observation duration and temporal aggregation, and sensitivity to sampling rate and deployment platform. The results establish the generality of A-GE across age-assurance policies and identify conditions that improve performance; the best-performing configuration is summarized at the end of this section.

\begin{figure}[t]
    \centering
    \begin{minipage}[t]{0.48\columnwidth}
        \centering
        \includegraphics[width=\linewidth]{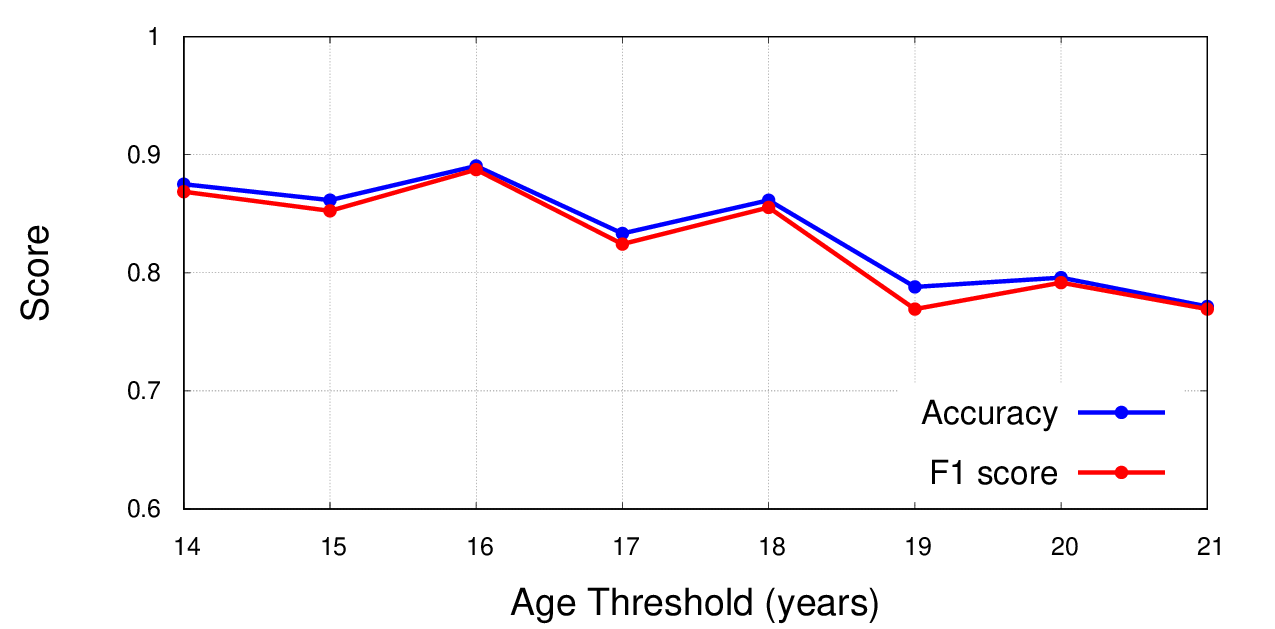}
        \caption{A-GE performance across minimum-age
        thresholds.}
        \label{fig:age-threshold}
    \end{minipage}
    \hfill
    \begin{minipage}[t]{0.48\columnwidth}
        \centering
        \includegraphics[width=\linewidth]{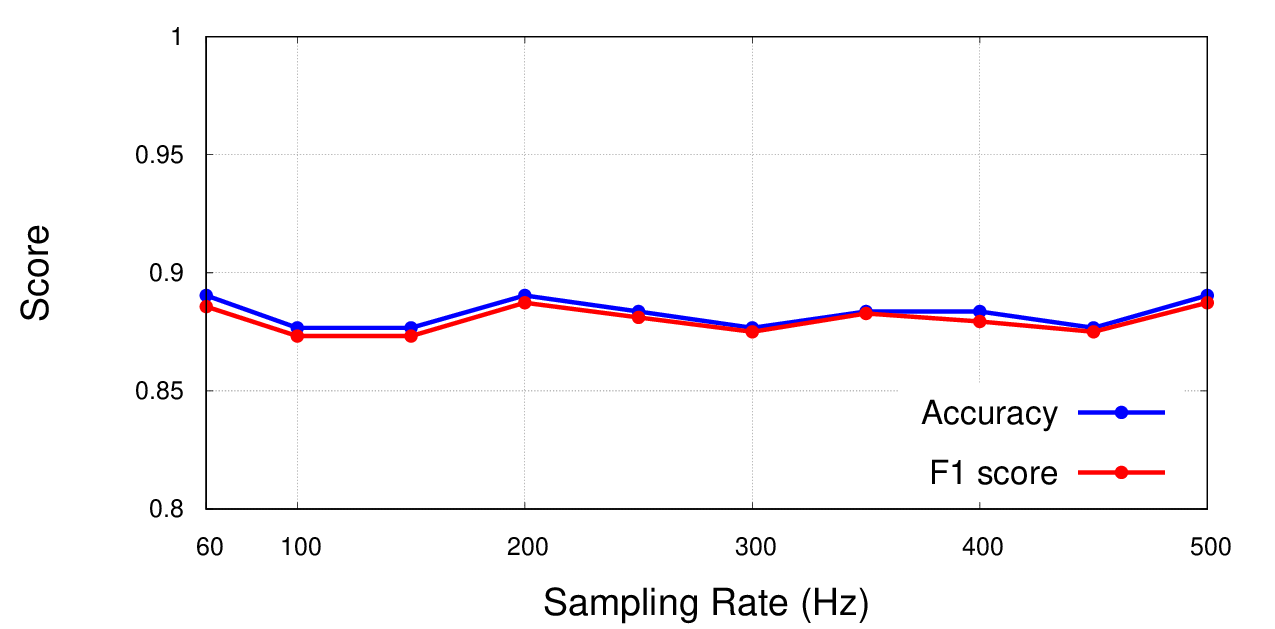}
        \caption{A-GE performance as a function of IMU sampling rate.}
        \label{fig:sampling-rate}
    \end{minipage}
\end{figure}

\textit{\textbf{Age-Threshold Generality.}}
A-GE is not intrinsically defined around a single age boundary. To examine whether the same sensing approach can support application-specific requirements, we train and evaluate age-assurance models using minimum-age thresholds from $14$ through $21$ years. For each threshold, the classification problem is defined independently, while all other experimental parameters are held constant. Figure~\ref{fig:age-threshold} shows the resulting performance across the evaluated thresholds. Performance varies across age boundaries, with the strongest performance observed at the $16$-year threshold and lower performance at some higher thresholds; for example, the $21$-year threshold achieves approximately $77\%$ accuracy under this fixed experimental setting. We do not interpret the peak at $16$ years as evidence of an intrinsic physiological optimum. The number and distribution of participants in the underage and adult classes vary with the threshold, and the resulting class composition can affect both model development and evaluation.

\textit{\textbf{Sampling-Rate Robustness.}}
Commodity devices expose substantially different inertial-sensing rates. Native mobile applications may provide measurements at rates of approximately $100$~Hz or higher, depending on the hardware, whereas browser-based interfaces may be limited to approximately $60$~Hz. A practical age-assurance system should therefore not depend on a narrow, high-frequency sensing configuration. To characterize sampling-rate robustness, we evaluate rates from $60$~Hz to $500$~Hz, using $60$~Hz and increments of $50$~Hz from $100$~Hz through $500$~Hz. Figure~\ref{fig:sampling-rate} shows the resulting performance for the $16$-year threshold and a $5$-second observation window. Across the evaluated rates, accuracy remains between $87.7\%$ and $89.0\%$, and F1 remains between $87.3\%$ and $88.7\%$, with the highest measured performance at $500$~Hz. The relatively small variation across sampling rates indicates that A-GE is robust to substantially different sensing configurations, with the ability to operate within browser-based sensor-access constraints. The same qualitative behavior is also observed across the evaluated minimum-age thresholds from $14$ through $21$ years.

\begin{figure}[t]
    \centering
    \begin{minipage}[t]{0.48\columnwidth}
        \centering
        \includegraphics[width=\linewidth]{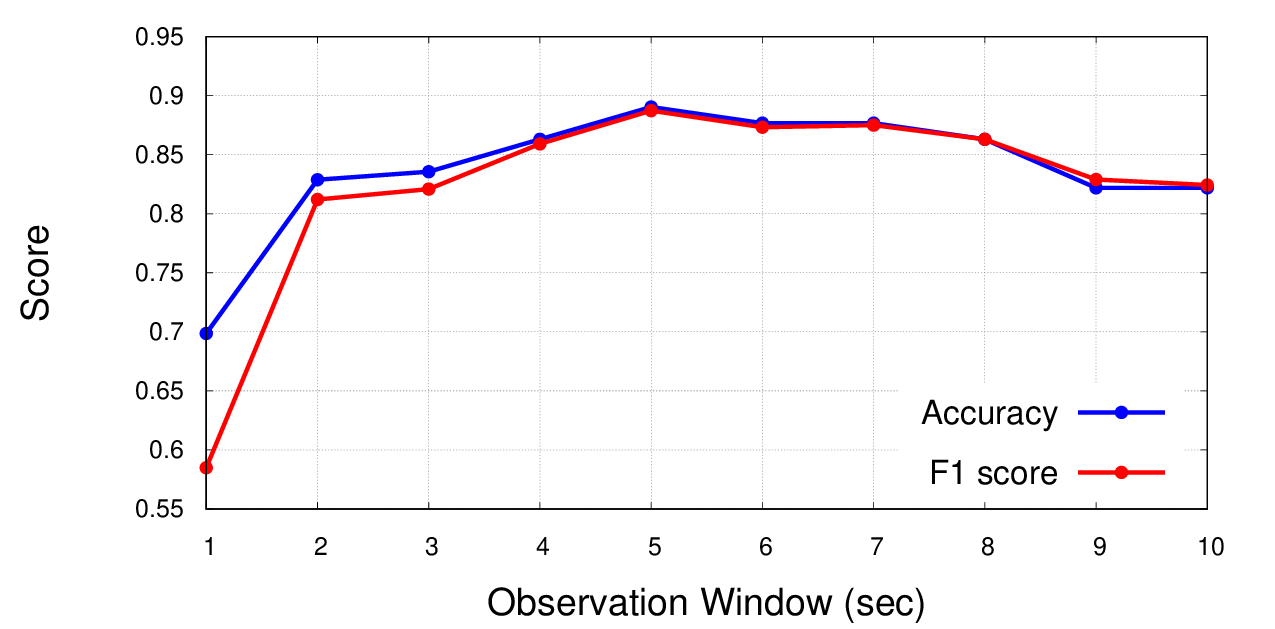}
        \caption{A-GE performance as a function of
        observation windows.}
        \label{fig:duration}
    \end{minipage}
    \hfill
    \begin{minipage}[t]{0.48\columnwidth}
        \centering
        \includegraphics[width=\linewidth]{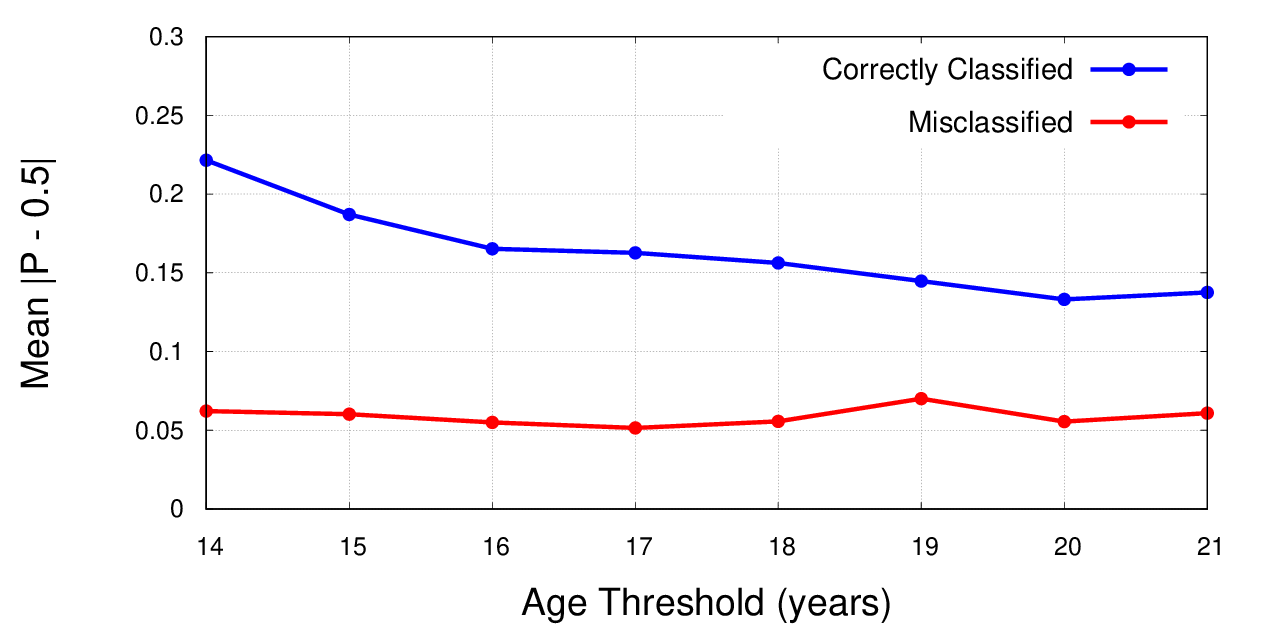}
        \caption{Mean decision margin, $|P-0.5|$, for correctly and
        incorrectly classified observations.}
        \label{fig:decision-margin}
    \end{minipage}
\end{figure}

\textit{\textbf{Observation Duration and Aggregation.}}
We next investigate two complementary approaches to improve the reliability of A-GE inference: increasing the duration of a single observation and aggregating decisions across multiple shorter observations. The former examines how much information is gained from longer sensor windows, while the latter evaluates whether repeated independent observations can improve decision reliability without requiring a longer continuous interaction.
%
Figure~\ref{fig:duration} shows that the performance improves substantially as the observation window increases from $1$ to approximately $5$ seconds.
%
Extending the observation interval beyond $5$ seconds does not necessarily translate into continued improvements. Considered together with the sampling-rate results shown earlier, which show only modest performance variation despite substantial changes in the number of acquired samples, these findings suggest that age-assurance performance depends more on the temporal characteristics of the observed motion than on the number of individual sensor samples alone.
These findings also motivate an alternative to extending a single observation window: combining multiple shorter observations. 

A-GE produces a probabilistic output rather than only a hard binary decision, allowing evidence from separate observations to be aggregated. For a predicted probability $P$ of belonging to the adult class, we define
the decision margin as
    $M = |P-0.5|$.
A small value of $M$ indicates that the prediction is close to the decision
boundary and is therefore comparatively ambiguous, whereas a larger value
indicates a more decisive prediction. Figure~\ref{fig:decision-margin}
shows the mean decision margin separately for correctly and incorrectly
classified observations across minimum-age thresholds from $14$ to $21$
years.
Across all evaluated thresholds, incorrectly classified observations exhibit
a consistently small mean decision margin, whereas
correctly classified observations have substantially larger margins. For
example, the mean margin for correctly classified observations decreases
from $0.2251$ at the $14$-year threshold to $0.1376$ at the $21$-year
threshold, while the corresponding margin for incorrectly classified
observations remains close to $0.05$. 
This behavior provides insight into the nature of individual classification errors. In particular, the substantially smaller margin of incorrectly classified observations indicates that, on average, incorrect decisions are associated with predictions close to the decision boundary rather than with highly separated predictions. This observation suggests that temporal aggregation can reduce the influence of ambiguous individual observations.

We therefore evaluate a practical aggregation strategy in which two independent observations are processed separately and their predicted adult probabilities are averaged before applying the decision threshold. Given two independent observations $X_1$ and $X_2$, we use their mean as the aggregated decision statistic:
    $P_{\mathrm{agg}}
    =
    \frac{1}{2} P(\mathrm{adult}\mid X_1)
        +
        P(\mathrm{adult}\mid X_2)$.
    \label{eq:prob-aggregation}
The resulting value is classified as adult when
$P_{\mathrm{agg}}\geq 0.5$.
Table~\ref{tab:temporal-aggregation} compares single-window and aggregated
observations. A single $3$-second observation achieves $83.6\%$ accuracy, while a single $6$-second observation provides $87.7\%$. In contrast, averaging the probabilities from two independent $3$-second observations increases performance to $91.1\%$. 
These results indicate that the temporal extent of the available evidence
and the manner in which it is aggregated are both important to practical
age assurance. 

\begin{table}[t]
\centering
\small

\begin{minipage}[t]{0.45\textwidth}
\centering
\caption{Effect of temporal observation and probability aggregation.}
\label{tab:temporal-aggregation}
\begin{tabular}{lcc}
\hline
\textbf{Observation strategy} & \textbf{Accuracy} & \textbf{F1} \\
\hline
$3$-s single window & 83.56\% & 82.09\% \\
$5$-s single window & 89.04\% & 88.73\% \\
$6$-s single window & 87.67\% & 87.32\% \\
$2\times3$-s probability aggregation & \textbf{91.10\%} & \textbf{90.78\%} \\
\hline
\end{tabular}
\end{minipage}
\hfill
\begin{minipage}[t]{0.45\textwidth}
\centering
\caption{A-GE performance by gender and age class.}
\label{tab:demographic-performance}
\begin{tabular}{lcc}
\hline
\textbf{Group} & \textbf{Accuracy (95\% CI)} & \textbf{F1} \\
\hline
Underage female & 96.88\% (84--99\%) & 87.32\% \\
Adult female    & 75.00\% (58--87\%) & 84.21\% \\
Underage male   & 87.50\% (72--95\%) & 88.89\% \\
Adult male      & 90.63\% (76--97\%) & 89.23\% \\
\hline
\end{tabular}
\end{minipage}

\end{table}

\textit{\textbf{Demographic Performance.}}
We next examine whether age-assurance performance varies across gender groups. This analysis is distinct from the attribute-inference experiments in Section~\ref{sec:privacy-analysis}: here, gender is used solely as a demographic stratification variable to evaluate the consistency of the intended age decision, rather than as an attribute that A-GE attempts to infer. To avoid class imbalance confounding the comparison, we construct a balanced evaluation subset containing equal numbers of female and male participants in both age classes, which is limited by 32 underage females. 
Table~\ref{tab:demographic-performance} reports accuracy, F1 score, and 95\% confidence intervals (CIs) for accuracy separately for each gender--age group. 
Performance varies across the four subgroups, 
whereas underage females achieve the highest.
The confidence intervals, however, are relatively wide because each subgroup contains only 32 participants.
Accordingly, the observed differences should be interpreted as measured variation rather than evidence of definitive demographic performance gaps.

\textit{\textbf{Deployment Overhead.}}
We evaluate A-GE's local inference efficiency on commodity mobile and web platforms: a Google Pixel 3a running Android 12, an iPhone 16 running iOS 26.6, and a browser-based implementation. We report inference latency, peak memory consumption, and CPU utilization; latency is measured from completion of a sensor window to production of the age decision, excluding sensor acquisition. Table~\ref{tab:deployment-overhead} shows that A-GE has low computational overhead across the evaluated platforms, supporting client-side execution without specialized hardware or remote computation. For a complementary view of classification performance, Table~\ref{tab:best-overall} summarizes the highest observed accuracy and F1 score for each age threshold, together with the corresponding configuration. These results characterize the observed performance envelope rather than a single configuration optimal for all thresholds; the preceding controlled experiments isolate the effects of individual factors.


\begin{table}[t]
\centering
\scriptsize

\begin{minipage}[t]{0.49\textwidth}
\centering
\caption{Computational overhead of client-side A-GE inference across platforms.}
\label{tab:deployment-overhead}
\begin{tabular}{llccc}
\hline
\textbf{Platform} &
\textbf{Rate} &
\textbf{Inference} &
\textbf{Memory} &
\textbf{CPU} \\
&
\textbf{(Hz)} &
\textbf{(ms)} &
\textbf{(MB)} &
\textbf{(\%)} \\
\hline
\multirow{4}{*}{Android}
    & 400 & 3.75 & 2.81 & 55.3 \\
    & 300 & 2.57 & 2.31 & 43.7 \\
    & 200 & 1.81 & 2.20 & 34.0 \\
    & 100 & 1.07 & 1.90 & 23.0 \\
\hline
iOS & 100 & 0.68 & 0.41 & 9.2 \\
\hline
\multirow{2}{*}{Web}
    & 60 (Android) & 0.65 & 1.77 & 17.6 \\
    & 60 (iOS) & 0.55 & 0.39 & 8.1 \\
\hline
\end{tabular}
\end{minipage}
\hfill
\begin{minipage}[t]{0.49\textwidth}
\centering
\caption{Best observed A-GE performance across evaluated experimental configurations.}
\label{tab:best-overall}
\begin{tabular}{ccccc}
\hline
\textbf{Age} &
\textbf{Accuracy} &
\textbf{F1} &
\textbf{Observation} &
\textbf{Sampling} \\
\textbf{threshold} &
&
\textbf{score} &
\textbf{strategy} &
\textbf{rate (Hz)} \\
\hline
14 & 95.2\% & 95.2\% & $1\times7$s & 450 \\
15 & 94.6\% & 94.7\% & $3\times3$s & 400 \\
16 & 93.8\% & 94.0\% & $2\times4$s & 300 \\
17 & 91.7\% & 91.8\% & $1\times9$s & 350 \\
18 & 91.6\% & 91.5\% & $1\times5$s & 350 \\
19 & 87.0\% & 87.2\% & $1\times9$s & 150 \\
20 & 86.2\% & 86.6\% & $2\times5$s & 350 \\
21 & 82.9\% & 83.5\% & $2\times5$s & 250 \\
\hline
\end{tabular}
\end{minipage}

\end{table}


\section{Conclusion}
\label{sec:conclusion}
A-GE explores privacy-preserving age assurance using fine-grained inertial measurements generated during ordinary interaction with commodity devices. It derives an application-defined age predicate locally from accelerometer and gyroscope measurements, without relying on identity documents, biometric images, persistent behavioral profiles, or a trusted age-credential issuer.
Our evaluation yields four main findings. \textit{First,} short observation windows can provide effective age assurance, while longer contiguous observations do not consistently improve performance. \textit{Second,} temporal aggregation can further improve performance. \textit{Third,} A-GE remains effective across substantial variation in IMU sampling rates. \textit{Finally,} the same sensing framework supports minimum-age thresholds from $14$ to $21$ years. Auxiliary analyses nevertheless show that IMU measurements encode information beyond the intended age property, reinforcing the importance of local processing and data minimization.
A-GE establishes the feasibility of the sensing and inference layer, but deployment requires end-to-end security guarantees. Future work will develop an architecture that protects sensor acquisition, feature extraction, inference, and decision making, and produces a cryptographically verifiable age-assurance certificate. This would complement local age inference with the integrity and verifiability needed for practical deployment.


%
%
%
\bibliographystyle{splncs04}
\bibliography{References}
%




\end{document}